\documentclass[journal,twoside,web]{ieeecolor}

\usepackage{generic}
\usepackage{cite}
\usepackage{amsmath,amssymb,amsfonts}

\usepackage{graphicx}
\usepackage{textcomp}
\def\BibTeX{{\rm B\kern-.05em{\sc i\kern-.025em b}\kern-.08em
		T\kern-.1667em\lower.7ex\hbox{E}\kern-.125emX}}
\usepackage[outdir=./figures/]{epstopdf}
\graphicspath{{./figures/}}

\usepackage[dvipsnames]{xcolor}
\usepackage{subfigure}
\usepackage{enumerate}

\usepackage{algorithm}
\usepackage[noend]{algpseudocode}
\usepackage{url}
\usepackage{cite}
\usepackage{hyperref}

\newcommand{\revised}[1]{{#1}}

\newcommand{\Vc}[1]{\mathbf{#1}}
\newcommand{\Bs}[1]{\boldsymbol{#1}}
\newcommand{\Cal}[1]{\mathcal{#1}}
\newcommand{\Rm}[1]{\mathrm{#1}}

\newcommand{\R}{\mathbb{R}}
\newcommand{\Z}{\mathbb{Z}}

\newcommand{\approxunion}{\mathcal{C}}

\newtheorem{theorem}{Theorem}
\newtheorem{lemma}{Lemma}

\newtheorem{assumption}{Assumption}
\newtheorem{example}{Example}

\newtheorem{remark}{Remark}

\begin{document}
		
	\bstctlcite{IEEEexample:BSTcontrol}
	
	\title{Safe Motion Planning against Multimodal Distributions based on a Scenario Approach}
	%	\author{First A. Author, \IEEEmembership{Fellow, IEEE}, Second B. Author, and Third C. Author, Jr., \IEEEmembership{Member, IEEE}
	\author{Heejin Ahn, Colin Chen, Ian M. Mitchell, and Maryam Kamgarpour
		\thanks{The authors were partially supported by National Science and Engineering Research Council of Canada (NSERC) Discovery Grants \#RGPIN-2017-04543 (Mitchell) and \#RGPAS-2020-00110 (Kamgarpour). }% <-this % stops a space
		\thanks{The authors are with the Departments of Computer Science (Ahn, Chen, Mitchell) and Electrical \& Computer Engineering (Ahn, Kamgarpour) at the University of British Columbia, Vancouver, BC, Canada. E-mail: {\tt\small hjahn@cs.ubc.ca}.}}
	\maketitle
	
	\thispagestyle{empty} % Removes the page number in the first page

	%%%%%%%%%%%%%%%%%%%%%%%%%%%%%%%%%%%%%%%%%%%%%%%%%%%%%%%%%%%%%%%%%%%%%%%%%%%%%%%%
	\begin{abstract}	
		%To plan safe motion of a vehicle, forecasting future motions of other vehicles is one of the required functionalities. Recent literature has reported successes in deep neural networks (DNNs) for such forecasting, yet the intrinsic problem of DNNs -- the fragility to uncertainties -- still remains. 
		We present the design of a motion planning algorithm that ensures safety for an autonomous vehicle. In particular, we consider a multimodal distribution over uncertainties; for example, the uncertain predictions of future trajectories of surrounding vehicles reflect discrete decisions, such as turning or going straight at intersections.
		We develop a computationally efficient, scenario-based approach that solves the motion planning problem with high confidence given a quantifiable number of samples from the multimodal distribution. Our approach is based on two preprocessing steps, which 1) separate the samples into distinct clusters and 2) compute a bounding polytope for each cluster.  Then, we rewrite the motion planning problem approximately as a mixed-integer problem using the polytopes. We demonstrate via simulation on the nuScenes dataset that our approach ensures safety with high probability in the presence of multimodal uncertainties, and is computationally more efficient and less conservative than a conventional scenario approach. 
	\end{abstract}
	
	%%%%%%%%%%%%%%%%%%%%%%%%%%%%%%%%%%%%%%%%%%%%%%%%%%%%%%%%%%%%%%%%%%%%%%%%%%%%%%%%
	\section{Introduction}
	\IEEEPARstart{S}{afe} motion planning in uncertain environments has been an active research area with consideration of various sources of uncertainties. Disturbances to dynamic models, e.g., wind disturbances, are the commonly studied source of uncertainties \cite{fisac_general_2019, wabersich_probabilistic_2021}. Uncertain specifications, e.g., uncertain obstacle locations, are another practical source \cite{carvalho_automated_2015, sessa_uncertainty_2018, lefkopoulos_using_2019}. Several past approaches assume a Gaussian distribution over uncertainties \cite{fisac_general_2019,lefkopoulos_using_2019, wabersich_probabilistic_2021} or represent the distribution via a finite number of samples without any assumption on it \cite{carvalho_automated_2015, sessa_uncertainty_2018}.
	
	%Recently, deep neural networks (DNNs) have been applied to and proven successful in safety-critical application domains, such as autonomous vehicles. However, DNNs are known to be fragile \cite{eykholt_robust_2018}. Failure can result in catastrophic incidents; for example, the Tesla \cite{tesla2016} fatal accident. In this paper, we present an approach to motion planning towards increasing system safety by accounting for uncertainties in the output of a DNN. 
	
\begin{figure}[tb]
		\centering
		\includegraphics[width=.9\linewidth]{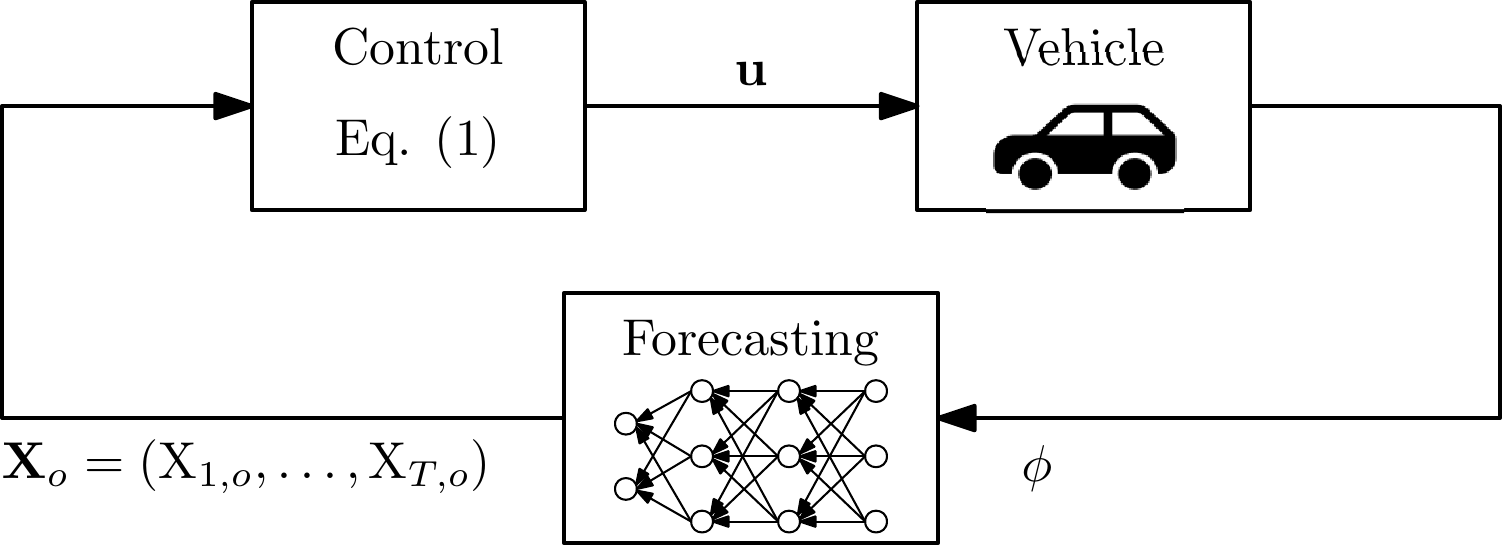}
		\caption{This paper presents a control design that ensures probabilistic safety against multimodal uncertainties. As a motivating example,  we solve the motion planning problem \eqref{eq:problem_motionplan_prob} based on other vehicles' predicted trajectories $\Vc{X}_{o}$ generated by a DNN.}
		\label{fig:closed-loop}
	\end{figure}

	{In this paper, we focus on a safe motion planning algorithm that can efficiently handle \textbf{multimodal} uncertainties.} We are motivated by an autonomous driving example, where future trajectories of other vehicles (OVs) are forecasted by a deep neural network (DNN), and a motion planner computes a trajectory of the ego vehicle (EV) that avoids any collision with the forecasts (See Fig.~\ref{fig:closed-loop}). 
	%Developing DNNs that learn a probability distribution over future trajectories of OVs has recently gained a lot of attention \cite{rhinehart_precog_2019, salzmann_trajectron_2021}.  
	%Unlike conventional DNNs, these DNNs also provide some information about the confidence on their predictions. 
	A noteworthy feature of most forecasting methods (e.g., \cite{salzmann_trajectron_2021}) is that the distribution over future trajectories is {multimodal}.
	%, i.e., the probability mass can be separated into distinct clusters. 
	In other words, predicted trajectories reflect different high-level decisions of OVs, such as braking/accelerating or going-straight/turning. Given prediction samples from the multimodal distribution, our goal is to design a motion planning algorithm that guarantees safety with high probability. {A similar problem is studied in \cite{ivanovic_mats_2021}, but focuses on the design of a forecasting method to facilitate the integration of forecasting into control.}
	
	Stochastic control with multimodal distributions is a more challenging problem than that with unimodal distributions. For example, if an uncertain parameter has a Gaussian distribution, a chance constraint can be reformulated as a convex constraint in terms of the quantile function \cite{calafiore_distributionally_2006}. However, with a mixture of two Gaussian distributions, such a convex reformulation is not possible due to the lack of a closed form for its quantile function. To our knowledge, there is no tractable way to solve general chance-constrained problems with multimodal distributions. In \cite{hu_chance_2018}, a branch-and-bound approach to solve chance-constrained problems with a mixture of Gaussian distributions is presented, but the approach is not tractable for large problems and thus not applicable to real-time implementation.

	We formulate the motion planning problem as a chance-constrained problem, and present an approximate yet efficient method to solve the problem when the uncertainty has a multimodal distribution. Specifically, we use the scenario approach \cite{esfahani_performance_2015} to characterize the number of samples $N$ required to solve the chance-constrained problem with high confidence. 
	\revised{We show that the conventional scenario approach \cite{esfahani_performance_2015} is only applicable to an approximation of the chance-constrained problem. Furthermore, it yields a conservative solution when the uncertainty has a multimodal distribution, and is computationally demanding as the number of mixed-integer constraints depends on the number of samples $N$.} Our approach to addressing these issues uses two preprocessing steps: First, we separate samples into distinct clusters to explicitly handle the multimodality, and next, we compute a bounding polytope for each cluster.
	%Because the conventional scenario problem has the computational complexity dependent on the number of samples $N$---which is often very large---we present an efficient method that eliminates such dependency. 
	With the preprocessing steps, we can rewrite the chance-constrained problem approximately as a deterministic problem, whose computational complexity does not depend on $N$. 
	%Although the computational complexity of the preprocessing steps depends on $N$, we formulate them as linear programs so that we can still achieve fast computation.
	Our approach is inspired by the preprocessing-based solution to chance-constrained problems \cite{margellos_road_2014}. We extend their technique to address the specific challenges of motion planning for autonomous driving: namely, non-convex collision-avoidance constraints and multimodal uncertainties.

	In summary, the contributions of this paper are:
	{
		\begin{itemize}
			%	\item We formulate a control problem that accounts for uncertainties in a DNN as a chance-constrained problem. 
			\item We show that applying the conventional scenario approach \cite{esfahani_performance_2015} to a motion planning problem yields conservative results when the uncertainty has a multimodal distribution and is computationally expensive.
			\item We present a computationally efficient method to solve the motion planning problem with high confidence.
			\item We test our approach with a state-of-the-art forecasting DNN on the nuScenes dataset \cite{caesar2020nuscenes} and validate that it ensures a desired level of safety, is computationally efficient, and is less conservative than the conventional scenario approach.
		\end{itemize}
	}
	
	%%%%%%%%%%%%%%%%%%%%%%%%%%%%%%%%%%%%%%%%%%%%%%%%%%%%%%%%%%%%%%%%%%%%%%%%%%%%%%%%
	%\subsection*{Notation}
	\noindent\textit{Notation:} We denote by $\mathbb{R}^n$ the set of $n$-dimensional real vectors, and $\mathbb{Z}, \mathbb{Z}_{a:b}$ by the sets of integers and integers bounded by $a\in \mathbb{Z}$ and $b\in \mathbb{Z}$, i.e., $\{a, a+1, \ldots, b\}$, respectively. Let $x_t \in \Cal{X} \subset \mathbb{R}^{n_x}, u_t\in \Cal{U} \subset \mathbb{R}^{n_u}, y_t\in \Cal{Y} \subset \mathbb{R}^{n_y}$ be the dynamical state, input, and output of the ego vehicle at time step $t$, respectively.  A sequence of inputs over $t\in \Z_{0:T-1}$ is denoted by the bold face $\Vc{u} := (u_0,\ldots, u_{T-1})$ in the space $\Bs{\Cal{U}} = \Cal{U}\times \cdots \times \Cal{U}$. Capital roman-type letters, such as $\Rm{X}$, indicate random variables.
	%, and bold capital letters indicate a sequence of random variables. For example, $\Vc{X}_{1:T}$ is a sequence of $\Rm{X}_t$ for $t=1,\ldots, T$. 
	%Given a set $A \in \R^n$, $(A)^c$ is the complement of $A$, i.e., $\R^n \setminus A$.
	
	%For two vectors $a, b\in \mathbb{R}^n$, we define the partial order relation $\leq$ such that $a\leq b \Leftrightarrow b-a \in \mathbb{R}^n_+$. , and bold capital letters indicate a sequence of random variables. For example, $\Vc{X}_{1:T}$ is a sequence of $\Rm{X}_t$ for $t=1,\ldots, T$, that is, $\Vc{X}_{1:T} = (\Rm{X}_1,\ldots, \Rm{X}_T)$.

	\section{Problem Setup}

	We aim to compute the input sequence $\Vc{u} \in \Bs{\Cal{U}} $ over a planning horizon of length $T$ that solves the following motion planning problem.
	\begin{subequations}\label{eq:problem_motionplan_prob}
		\begin{align}
		&\min_{\Vc{u}\in \Bs{\Cal{U}}} && J(\Vc{u})\\
		&\text{s.t.} && {x}_{t+1} = f_t(x_t,u_t), \quad t\in\Z_{0:T-1},  \label{opt1:model}\\
		& && y_t = h_t(x_t), \quad t\in \Z_{1:T} , \label{opt1:output} \\
		& && \Pr\left(\bigwedge_{t=1}^T \bigwedge_{o=1}^O {y}_t \notin  \Cal{O}(\Rm{X}_{t,o}) \right) \geq 1-\epsilon.\label{opt1:chance-constraint}
		\end{align}
	\end{subequations}
	Here, $J:\Bs{\Cal{U}}\rightarrow \R$ is a convex objective function that guides the EV toward goal states, \eqref{opt1:model} is the discrete-time dynamical model of the EV, \eqref{opt1:output} is the output, and \eqref{opt1:chance-constraint} is a chance constraint that encodes the avoidance of collisions. In constraint \eqref{opt1:chance-constraint}, $\Rm{X}_{t,o}\in \R^{n_{\Rm{X}}}$ is the predicted state of OV $o$ at time $t$, and $\Cal{O}(\Rm{X}_{t,o})\subset \mathbb{R}^{n_y}$ is an obstacle set that the EV's output $y_t$ should avoid. Since $\Rm{X}_{t,o}$ is a random variable, we impose the probability of non-collision to be no smaller than $1-\epsilon$, where $\epsilon\in (0,1)$ indicates a risk level.
	The number of OVs around the EV is $O$. 
	
	{Consider that the dynamical model is linear time-varying, that is, at each time step $t$, $f_t$ and $h_t$ in \eqref{opt1:model} and \eqref{opt1:output} are linear functions of $x_t$ and $u_t$.} Also, let $\Cal{O}(\Rm{X}_{t,o})$ be a polytope expressed by 
	\begin{align}\label{eq:obstable-polytope}
	\Cal{O}(\Rm{X}_{t,o}) := \{y \in \R^{n_y}: A(\Rm{X}_{t,o}) y < b(\Rm{X}_{t,o})\},
	\end{align}
	which is the intersection of $L$ halfspaces defined by nonlinear continuous functions $A: \R^{n_\Rm{X}}\rightarrow \R^{L \times n_y}$ and $b: \R^{n_\Rm{X}}\rightarrow \R^{L}$. To encode constraints of polytopic obstacle avoidance, it is common to use the big-$M$ method \cite{bemporad1999control} where binary variables are introduced to indicate whether the inequalities $A(\Rm{X}_{t,o}) y < b(\Rm{X}_{t,o})$ in \eqref{eq:obstable-polytope} are satisfied.
	\revised{Specifically, we can introduce a binary variable $z_{j,t,o}(\Rm{X}_{t,o}) \in \{0,1\}$ as a function of $\Rm{X}_{t,o}$ and rewrite $y_t\notin \Cal{O}\left(\Rm{X}_{t,o}\right)$ as
		\begin{align}\label{eq:big-M}
		A_j\left(\Rm{X}_{t,o} \right) y_t + M \left(1-z_{j,t,o}(\Rm{X}_{t,o})\right) \geq b_j \left(\Rm{X}_{t,o}\right), ~\forall j \in \Z_{1:L},
		\end{align}
		and $\sum_{j=1}^{L} z_{j,t,o}(\Rm{X}_{t,o}) \geq 1,$} where $M \in \R$ is a sufficiently large number, and $A_j$ and $b_j$ are the $j$-th row of $A$ and $b$, respectively. Here, $z_{j,t,o}(\Rm{X}_{t,o}) = 1$ indicates that one of the inequalities in \eqref{eq:obstable-polytope} is violated (i.e., $A_j(\Rm{X}_{t,o}) y_t \geq b_j(\Rm{X}_{t,o})$), which implies that output $y_t$ is outside $\Cal{O}(\Rm{X}_{t,o})$.

	%\begin{remark}
	%s an individual chance constraint at planning time step $t$. We can instead enforce the joint chance constraint $\Pr(\bigwedge_{t=1}^T {y}_t \notin \Cal{O}(\Rm{Y}_t) ) \geq 1-\epsilon$, and approximate it as individual chance constraints by applying Boole's inequality \cite{lefkopoulos_using_2019}. 	
	%\end{remark}

	Next, we explain forecasting algorithms that generate $\Vc{X} = (\Rm{X}_{t,o}: t\in \Z_{1:T}, o\in \Z_{1: O})$, and discuss our approach to solving problem \eqref{eq:problem_motionplan_prob} based on the forecasts.
	
	\subsection{Forecasting}
	
	%\begin{figure}[thb]
	%	\centering
	%	\includegraphics[width=0.6\columnwidth]{forecasting-result.jpg}
	%	\caption{\revised{(The figure will be replaced.)}}
	%	\label{fig:DNNoutput}
	%\end{figure}
	Forecasting algorithms generate future trajectories of OVs based on context $\phi$, which is often a motion history of each OV and current road structure perceived by sensors, e.g., cameras and lidars. Given a training dataset of ground-truth future trajectories and contexts, i.e., $(\Vc{X}^i, \phi^i)$, the goal of recent forecasting DNNs (e.g., \cite{salzmann_trajectron_2021}) is to learn the conditional probability distribution $p(\Vc{X}|\phi)$, so that at test time, they can extract arbitrarily many predictions from it.
	
	Since we focus on the design of a motion planner, we assume that a forecasting DNN is well-trained with a sufficient amount of training data and accurately predicts possible future trajectories of OVs. 
	This is to avoid out-of-distribution scenarios (i.e., scenarios that have not been observed during training), which may cause incorrect forecasts. The relaxation of this assumption leads to a different research problem actively studied in machine learning communities \cite{nalisnick_deep_2018}, and remains as future work.

	%This generative modeling of DNNs enables the computation of confidence levels on their forecasts (e.g., small variances indicate confident forecasts).
	%In this paper, we focus on the prediction of other vehicles' future trajectories, but the control design presented here can be applied to other traffic participants, such as pedestrians. 
	%Mixture models, conditional generative models, and Markov decision processes based on reward map learning have been studied. 

	%\eqref{opt1:chance-constraint} changes to 
	%\begin{align}
	%\inf_{q  \in \Cal{P}} q({y}_t \notin \Cal{O}(\Rm{X}_{t,o}) ) \geq 1-\epsilon, && o\in \Z_{1:M}, t\in \Z_{1:T}.
	%\end{align}
	%The set of distributions 
	%\begin{align}
	%\Cal{P} &:= \{q \in \Cal{M}:  (H(p, q) - H(\eta)) / (Tm) \leq \varepsilon\},\\
	%\end{align}
	%where $H(p,q)$ is the cross entropy between two distributions $p$ and $q$, that is,
	%\begin{align}
	%H(p,q) & = -\mathbb{E}_{x\sim p} \log q(x),
	%\end{align}
	%and $\eta = \Cal{N}(0, 0.01 I)$.
	
	\subsection{Scenario Approach}
	A common approach to chance-constrained problems is the scenario approach \cite{calafiore_uncertain_2005}, which represents the uncertainty distribution via a finite number of samples. When the distribution is modeled by a DNN, we can query the DNN to generate samples from the distribution. 
	\revised{While \cite{calafiore_uncertain_2005} is applicable only to convex problems, the work \cite{esfahani_performance_2015} extends it to mixed-integer problems. However, the reformulation \eqref{eq:big-M} does not fit in the formulation from \cite{esfahani_performance_2015} because binary decision variables in \eqref{eq:big-M} depend on the uncertain parameter $\Rm{X}_{t,o}$. Thus we instead consider the following approximation of problem \eqref{eq:problem_motionplan_prob}.}
	\begin{subequations}\label{eq:problem_mp_explicit}
		\begin{align}
		&\min_{\Vc{u}\in \Bs{\Cal{U}}, \Vc{z}\in \Cal{Z}}  ~J(\Vc{u})\\
		&\text{s.t.} \hspace{1 cm} \eqref{opt1:model}, \eqref{opt1:output} \\
		& \Pr\left(\bigwedge_{j, t, o} A_j\left(\Rm{X}_{t,o} \right) y_t + M \left(1-z_{j,t,o}\right)\geq b_j \left(\Rm{X}_{t,o}\right)\right) \geq 1-\epsilon, \label{opt:mp_bigM}
		\end{align}
	\end{subequations}
	where \revised{$\Vc{z}$ is a sequence of binary variables $z_{j,t,o}\in \{0,1\}$ for $j\in \Z_{1: L}, t\in \Z_{1:T}, o\in \Z_{1:O}$, which are independent} \revised{of $\Rm{X}_{t,o}$ in contrast to \eqref{eq:big-M}}, and $\Cal{Z} := \{\Vc{z}\in \{0,1\}^{L TO}: \sum_{j=1}^L z_{j,t,o} \geq 1\}$. 
	The approximate problem \eqref{eq:problem_mp_explicit} involves $n_c = Tn_u$ continuous and $n_b = L TO$ binary variables.

	Given $N$ independent and identically distributed samples $\Rm{X}_{t,o}^{(i)}$ for $t\in \Z_{1:T}$ and $o\in \Z_{1:O}$, {we evaluate the constraints in problem \eqref{eq:problem_mp_explicit} at the samples} as follows:
	\begin{subequations}\label{eq:problem_scenario}
		\begin{align}
		&\min_{\Vc{u} \in \Bs{\Cal{U}}, \Vc{z} \in \Cal{Z}} && J(\Vc{u})\\
		&\text{s.t.} && \eqref{opt1:model}, \eqref{opt1:output}\\
		& &&A_j\left(\Rm{X}_{t,o}^{(i)} \right)^\intercal y_t + M \left(1-z_{j,t,o}\right) \geq b_j \left(\Rm{X}_{t,o}^{(i)}\right), \notag\\
		& && \hspace{0.4 cm}  j\in \Z_{1: L}, t \in \Z_{1:T}, o \in \Z_{1:O}, i \in \Z_{1:N}.\label{opt2:chance-constraint}
		\end{align}
	\end{subequations}
	We can apply the main result of \cite{esfahani_performance_2015}.
	%Problem \eqref{eq:problem_scenario} is \textbf{a mixed-integer program}. 
	
	% quantify the number of samples $N$ required to guarantee the optimal solution of scenario problem \eqref{eq:problem_scenario} to be feasible to chance-constrained problem \eqref{eq:problem_motionplan_prob}. , which is applicable only for continuous decision variables,
	
	%Note that constraint \eqref{opt2:chance-constraint} follows the reformulation \eqref{eq:big-M}, but it is not equivalent to 
	%\begin{align}\label{eq:sampled_obstacle_set}
	%y_t\notin \Cal{O}\left(\Rm{X}_{t,o}^{(i)}\right),~ t \in \Z_{1:T}, o \in \Z_{1:O}, i \in \Z_{1:N}.
	%\end{align} Specifically, \eqref{opt2:chance-constraint} implies \eqref{eq:sampled_obstacle_set}, but the conserve is not true. This is because \eqref{opt2:chance-constraint} does not have a sufficient number of binary variables to encode \eqref{eq:sampled_obstacle_set} as $\Vc{z}$ in \eqref{opt2:chance-constraint} does not depend on sample index $i$. Since we formulate scenario problem \eqref{eq:problem_scenario} to have the same number of decision variables as original problem \eqref{eq:problem_motionplan_prob} based on the scenario-based formulation given in \cite{esfahani_performance_2015}, we can apply the following generalization result. 
	
	\begin{lemma}[\cite{esfahani_performance_2015}]\label{lemma:scenario}
		Given a risk level $\epsilon\in (0,1)$ and a confidence bound $\beta \in (0,1)$, let $N$ satisfy
		\begin{align}
		2^{n_b} \sum_{i=0}^{n_c-1} {N \choose i} \epsilon^i (1-\epsilon)^{N-i} \leq \beta,
		\label{eq:N-condition}
		\end{align}
		where $n_c = Tn_u$ and $n_b = L TO$.
		{Then, the optimal solution of \eqref{eq:problem_scenario} is feasible for \eqref{eq:problem_mp_explicit}} with probability at least $1-\beta$.
	\end{lemma}
	
	%	Lemma~\ref{lemma:scenario} provides the lower bound on the number of samples $N$ that makes the solution of \eqref{eq:problem_scenario} feasible to \eqref{eq:problem_mp_explicit} with a desired confidence level of $1-\beta$. 
	
	%	Specifically, it is shown in \cite{alamo_randomized_2015} that $N$ satisfies \eqref{eq:N-condition} if 
	%	\begin{align}
	%	N \geq \frac{1.59}{\epsilon} \left( \ln \frac{2^{n_b}}{\beta} + n_c - 1\right).
	%	\label{eq:N-lower-bound}
	%	\end{align}
	%	The required number of samples $N$ increases linearly with the number of decision variables $n_c$ and $n_b$ and with the inverse of risk level $\epsilon$, and logarithmically with the inverse of confidence level $\beta$. 
	%Thus, the number of samples $N$ increases more sensitively to $n_c, n_b,$ and $\epsilon$ than $\beta$.

	\subsection{Drawbacks of Scenario Problem \eqref{eq:problem_scenario}}
	Although problem \eqref{eq:problem_scenario} is readily solvable by available mixed-integer program solvers, it has two drawbacks. First, the number of mixed-integer constraints \eqref{opt2:chance-constraint} is $LTON$. This means that \eqref{eq:problem_scenario} involves a large number of mixed-integer constraints because the lower bound of $N$ that satisfies \eqref{eq:N-condition} is usually large. For example, in one of the simulations in Section~\ref{sec:simulation}, $N = 1706$ and solving \eqref{eq:problem_scenario} takes about $4\,$s, which indicates that this approach may not be suitable for real-time applications. Second, scenario problem \eqref{eq:problem_scenario} tends to be conservative, especially when $\Rm{X}_{t,o}$ has a multimodal distribution. We illustrate this aspect in the following example.
	
	\begin{figure}[t]
		\centering
		\subfigure[Example~\ref{ex:scenario-approach}]{
			\includegraphics[width=0.46\columnwidth]{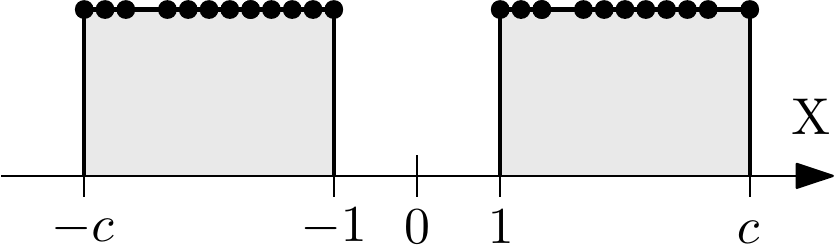}
		}
		\subfigure[Example~\ref{ex:preprocessing}]{
			\includegraphics[width=0.46\columnwidth]{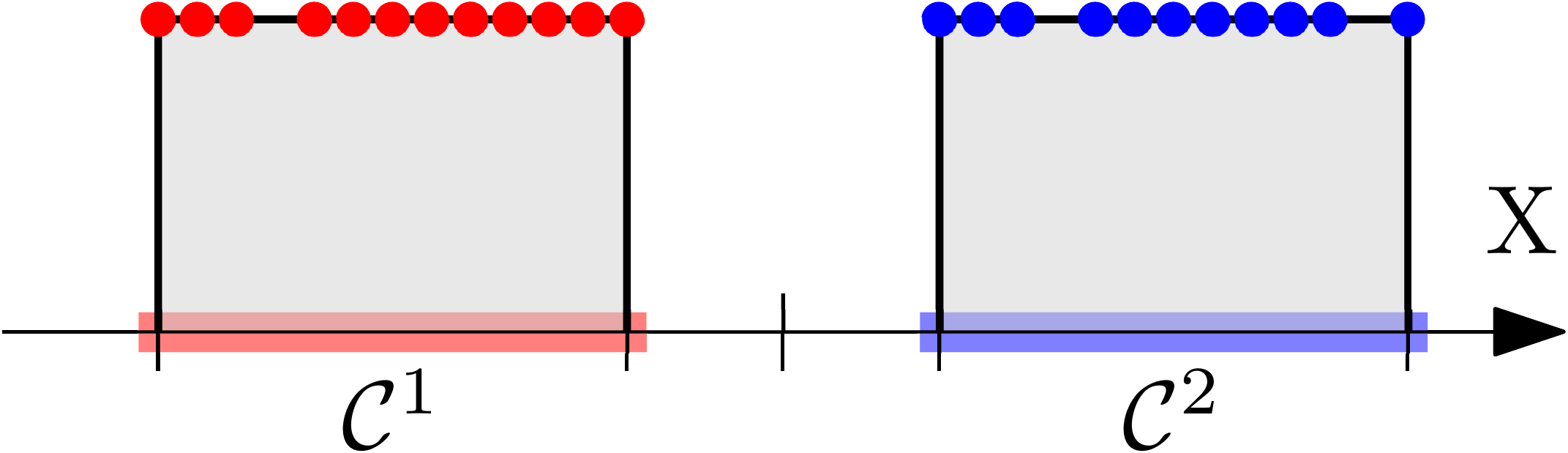}
		}
		
		\caption{(a) In Example~\ref{ex:scenario-approach}, we solve scenario problem \eqref{eq:problem_scenario} based on samples drawn from a mixture of two uniform distributions. The dots are a subset of the samples. (b) In Example~\ref{ex:preprocessing}, we separate the samples into two clusters (red and blue) and compute the set $\Cal{C}^k$ that contains the obstacle sets of samples in cluster $k$.}
		\label{fig:scenario-example}
	\end{figure}
	
	\begin{example}\label{ex:scenario-approach}
		Consider a simple one-dimensional example where $u = y\in \R$ and $\Rm{X}_{t,o}\in \R$. We let $T=1, O=1$ and omit the subscripts $t,o$ in the random variable $\Rm{X}_{t,o}$. Suppose that the distribution of $\Rm{X}$ takes the form of a mixture of two uniform distributions as depicted in Fig.~\ref{fig:scenario-example}(a), from which we draw $N$ samples. Suppose that the maximum and minimum values of the samples are $c$ and $-c$, respectively, for some positive constant $c\in \mathbb{R}$. Let  $\Cal{O}(\Rm{X}) = (\Rm{X}-0.1, \Rm{X}+0.1)\subset \R$ and thus $L =2$. The scenario problem \eqref{eq:problem_scenario} that minimizes $|y|$ then becomes
		\begin{subequations}\label{eq:problem_example}
			\begin{align}
			&\min_{y, \Vc{z}\in \Cal{Z}}  && |y|\\
			&\text{s.t.} && y + M \left(1-z_1\right) \geq \Rm{X}^{(i)} + 0.1, \quad i \in \Z_{1:N},\\
			& &&-y + M \left(1-z_2\right) \geq -\Rm{X}^{(i)} + 0.1, i \in \Z_{1:N},
			\end{align}
		\end{subequations}
		where $\Cal{Z} = \{(z_1, z_2) \in \{0,1\}^2:  z_1 + z_2 \geq 1\}$.
		By Lemma~\ref{lemma:scenario}, if $N \geq 190.6$, the optimal solution of \eqref{eq:problem_example} guarantees {\eqref{opt:mp_bigM} with $\epsilon = 0.05$} with probability at least $1-0.01$. 
		Note that $y=0$ avoids entering all the obstacle sets, i.e., $y\notin \Cal{O}(\Rm{X}^{(i)})$ for all $i$, as no sample lies in between $-1$ and $1$. However, in problem \eqref{eq:problem_example}, $y=0$ is not a feasible solution because there is no associated feasible set of binary variables. Instead, the optimal cost of \eqref{eq:problem_example} is $c+0.1$, which can be arbitrarily large.
	\end{example}
	
	%	\red{I think actually the Remark 1 comment could be added here and this way, you can address two of the comments of that reviewer: 1) on formulating scenario diretly on 1; 2) potential benefits for this problem. at the same time highlighting the large number of binaries and the fact that we want a priori... }
	%	\revised{
	%		\begin{remark}
	%			Instead of formulating problem \eqref{eq:problem_scenario} as a scenario counterpart of problem~\eqref{eq:problem_mp_explicit}, it is also possible to formulate a scenario counterpart of problem~\eqref{eq:problem_motionplan_prob}.
	%			%, which replaces \eqref{opt1:chance-constraint} with $\bigwedge \bigwedge y_t\notin \Cal{O}(\Rm{X}_{t,o}^{(i)}), \forall i\in \Z_{1:N}$. 
	%			However, we have chosen to work with problem~\eqref{eq:problem_scenario} because then we can apply the generalization result in \cite{esfahani_performance_2015} (i.e., Lemma~\ref{lemma:scenario}). The scenario counterpart of problem~\eqref{eq:problem_motionplan_prob}, on the other hand, does not fit in the scenario-based formulation in \cite{esfahani_performance_2015}. As an alternative, we can apply the result of a recent scenario theory \cite{campi_general_2018}. However, the result of \cite{campi_general_2018} provides an \textit{a-posteriori} bound on the risk level $\epsilon$, which means that we can only evaluate $\epsilon$ after solving the scenario problem. Since our goal is to ensure safety at an \textit{a-priori} risk level $\epsilon$, we consider scenario problem \eqref{eq:problem_scenario}.
	%		\end{remark}
	%	}
	
	Example~\ref{ex:scenario-approach} illustrates that scenario-based formulation \eqref{eq:problem_scenario} may result in a conservative solution due to an insufficient number of binary variables when uncertain parameters have a multimodal distribution. If we introduce an additional set of binary variables for each mode of the distribution, we can reduce the conservativeness. This is the intuitive idea behind our approach, which we will explain in the next section.

	%In the following section, we present an approach that overcomes the drawbacks of conservativeness and computational complexity.
	
	%%%%%%%%%%%%%%%%%%%%%%%%%%%%%%%%%%%%%%%%%%%%%%%%%%%%%%%%%%%%%%%%%%%%%%%%%%%%%%%%
	
	\section{Solution Approach}\label{sec:overapproximating}
	
	Our approach to solving problem \eqref{eq:problem_motionplan_prob} is to:
	\begin{enumerate}[A)]
		\item Separate samples into $K$ clusters where $K$ is the number of modes in the distribution.
		\item Compute a set $\approxunion^k$ that contains the obstacle sets in each cluster $k$.
		\item Formulate a deterministic motion planning problem given $\bigcup_{k=1}^K \approxunion^k$. 
	\end{enumerate}
	In step A, our approach explicitly handles $K$ modes of the distribution to reduce the conservativeness. We formulate step B as a computationally efficient linear program, and use the result to make the computation of the more challenging, mixed-integer motion planning problem in step C independent of $N$. 
	%Overall the resulting approach is more computationally efficient than the basic scenario approach. 
	In this section, we explain the three steps in detail.
	
	%By applying the generalization result of the scenario approach (i.e., Lemma~\ref{lemma:scenario}), we can quantify the lower bound of the number of samples $N$ required to ensure that $\approxunion$ contains any possible realization of $\Cal{O}(\Rm{X}_{t,o})$ with high probability. This in turn implies that the solution of the robust problem is a feasible solution to \eqref{eq:problem_motionplan_prob} with high probability. 
	
	\subsection{Clustering}\label{section:clustering}
	{There are several approaches to the clustering step. Given the number of modes $K_o$ of the probability distribution for OV $o$, one approach is to use the $k$-means clustering method \cite{bishop2006pattern} to separate samples into $K_o$ clusters. Note that modes are often known in many driving scenarios. For example, at intersections, vehicles can either go straight or turn, in which case clustering amounts to determining whether $\Rm{X}_{t,o}$ corresponds to going straight or turning. Instead of running the clustering algorithm for each $t\in \Z_{1:T}$, the clustering algorithm can be run for samples $\Rm{X}_{T,o}^{(i)}, \,i\in \Z_{1:N}$ at the last prediction time step $T$. This is due to the observation that in almost all cases modes exhibit the greatest separation at the longest prediction time $t$.  An alternative approach is to directly use the data provided from forecasting neural networks. In particular, such networks  commonly use latent variables to encode driving modes (e.g., \cite{salzmann_trajectron_2021}), and can  provide a forecast $(\Rm{X}_{1,o}, \ldots, \Rm{X}_{T,o})$ together with the mode. In either case, for each OV $o$, the clustering algorithm partitions samples $\Rm{X}_{t,o}^{(i)}, i\in\Z_{1:N}$ into $K_o$ clusters: $\Rm{X}_{t,o}^{(i)} \text{ for } i\in I_o^k, ~k\in \Z_{1:K_o}$, where $I_o^k$ is the set of sample indexes in cluster $k$, $\bigcap_k I_o^k = \emptyset$, $ \bigcup_{k} I_o^k = \Z_{1:N}$, and is the same for all time steps $t=1,\ldots, T$.}
	
	{Based on the above discussion, we assume that the clustering is a predetermined process.}
	{
		\begin{assumption}\label{assume:clustering}
			Clustering is done via a function $g: \R^{n_\Rm{X}} \rightarrow \Z_{1:K_o}$ that assigns each realization $\Rm{X}_{t,o} \in \R^{n_\Rm{X}}$ to a corresponding cluster $g(\Rm{X}_{t,o})\in \Z_{1:K_o}$.
		\end{assumption}
	}
	{We demonstrate the two ways mentioned above of defining $g$ in Section~\ref{sec:simulation}.}
	%	\red{ Assumption~\ref{assume:clustering} is reasonable in various practical cases.
	%	One interpretation of Assumption~\ref{assume:clustering} is that a clustering algorithm determines decision boundaries (thus, a function $g$) based on a training dataset, which can then be used to perform classification of testing data. 
	
	% Removed the probability information as it seems out of place..
	% One useful information that often comes with clustering is the probability of being classified to a cluster, i.e., $p(g(\Rm{X}_{t,o}))$. 
	%	 that the number of modes $K_o$ is known in advance. This is possible in various driving scenarios due to road regulations; for example, vehicles can either follow the current lane or change lanes, in which case $K_o=2$. When the number of modes is not exactly known, an approximation of the number can still ensure collision avoidance at the cost of additional computational complexity or conservativeness.

	\subsection{Polytopic Approximation}\label{section:overapproximation}
	If a set $\approxunion_{t,o}$ satisfies $\bigcup_{i=1}^N \Cal{O}(\Rm{X}_{t,o}^{(i)})\subseteq \mathcal{C}_{t,o},$
	then constraint $y_t \notin \Cal{O}(\Rm{X}_{t,o}^{(i)}), \forall i\in \Z_{1:N}$ is implied by $y_t \notin \Cal{C}_{t,o}$.
	%Thus, we aim to obtain a tight overapproximation $\approxunion_{t,o}$. 
	For the tractable computation of the collision avoidance constraint $y_t \notin \Cal{C}_{t,o}$, we represent $\approxunion_{t,o}\subset \mathbb{R}^{n_y}$ by the  union of polytopes 
	$\bigcup_{k=1}^{K_o} \approxunion_{t,o}^k:= \bigcup_{k=1}^{K_o} \left\{ y\in \mathbb{R}^{n_y}:  \bar{A}_{t,o}^k y \leq \bar{b}_{t,o}^k \right\}$,  
	where $\approxunion_{t,o}^k$ is a polytope that contains the obstacle sets in cluster $k$, $\bar{A}_{t,o}^k \in  \mathbb{R}^{L_{\approxunion}\times n_y}, \bar{b}_{t,o}^k\in \mathbb{R}^{L_{\approxunion}}$, and $L_{\approxunion}$ is the number of halfspaces. 
	%	Since $\Rm{X}_{t,o}^{(i)}$ in each cluster are in the vicinity of each other and $\Cal{O}(\Rm X_{t,o}^{(i)})$ is continuous in $\Rm{X}_{t,o}^{(i)}$, the obstacle sets $\Cal{O}(\Rm{X}_{t,o}^{(i)})$ in each cluster are also close to each other. Using this fact, we can obtain a tight overapproximation $\Cal{C}_{t,o}^k$. 

	%	\red{I still find that couple of clarificaitons could be added here: 
	%1) why predeterming $A_t$ is reasonable ? how we do it; 
	%2) why the heuristic cost is a good "heuristic" approximation? in which sense.}
	To efficiently obtain $\approxunion_{t,o}^k$, we predetermine $\bar{A}_{t,o}^k$ and let $\bar{b}_{t,o}^k$ be the only decision variable. For example, we set $\bar{A}_{t,o}^k$ to be $A_j(\bar{\Rm{X}}_{t,o}^k)$, the same matrix as in the obstacle set, evaluated at the element-wise average $\bar{\Rm{X}}_{t,o}^k := \frac{1}{|I_o^k|} \sum_{i\in I_o^k} \Rm{X}_{t,o}^{(i)}$.
	Moreover, we use the vertexes of $\Cal{O}(\Rm{X}_{t,o}^{(i)})$ to significantly speed up the computation. For example, the vertexes are easily computable when we model the obstacle set as a rectangle of the OV width and length, rotated with the OV yaw angle. Let $\Cal{V}_{t,o}^k$ be the set of the vertexes of $\Cal{O}(\Rm{X}_{t,o}^{(i)})$ for $i\in I_o^k$. Then, the overapproximation $\Cal{C}_{t,o}^k$ should contain all the vertexes in $\Cal{V}_{t,o}^k$ due to convexity. The following program computes {$\Bs{\approxunion}_{o}^k = (\approxunion_{1,o}^k, \ldots, \approxunion_{T,o}^k)$ for each cluster $k$ of OV $o$.}{\begin{subequations}\label{eq:problem-LP}
			\begin{align}
			&\min_{\bar{b}_{t,o}^k, \forall t} &&  \sum_{t=1}^T \bar{J}(\bar{b}_{t,o}^k)\\
			&\text{s.t.} && \bar{A}_{t,o}^k v \leq \bar{b}_{t,o}^k,  && v\in \Cal{V}_{t,o}^k, t\in \Z_{1:T}.\label{eq:LP-constraint}
			\end{align}
		\end{subequations}
	}
	We use $\bar{J}(\bar{b}) = \sum_j \bar{b}_j$ as a heuristic to yield a tight approximation because a larger $\bar{J}(\bar{b})$ then indicates a larger volume of polytope $\{y: \bar{A}y\leq \bar{b}\}$ when $\bar{A}$ is fixed. {Problem~\eqref{eq:problem-LP} consists of $L_{\approxunion} T$ decision variables, and $L_{\approxunion} \sum_t | \Cal{V}_{t,o}^k |$ constraints.} Although the number of constraints increases with the number of samples $N$ since $|\Cal{V}_{t,o}^k| = O(LN)$, it can be efficiently solved by linear program solvers, or by applying an element-wise maximum operator to $\bar A^{k}_{t,o} v$ in \eqref{eq:LP-constraint} over all $v$ because with our choice of the cost function, the optimal solution is $\bar{b}^k_{t,o} = \max_{v\in \Cal{V}^k_{t,o}}(\bar A^{k}_{t,o} v)$. 

	\subsection{Mixed-integer Motion Planning Formulation}\label{sec:motion_plan_mip}
	Using the overapproximation of the obstacle sets, we formulate a deterministic motion planning problem.
	\begin{subequations}\label{eq:problem_motionplan_overapprox}
		\begin{align}
		&\min_{\Vc{u}\in \Bs{\Cal{U}}} && J(\Vc{u})\\
		&\text{s.t.} && \eqref{opt1:model}, \eqref{opt1:output},\\
		& &&  {y}_{t} \notin \approxunion_{t,o}^k  , && k\in \Z_{1:K_o}, t \in \Z_{1:T},o\in \Z_{1:O}. \label{mp1:collision-avoidance}
		\end{align}
	\end{subequations}
	Problem \eqref{eq:problem_motionplan_overapprox} can be written as a mixed-integer program.
	%For all OV, problem~\eqref{eq:problem-LP} contains $ L_\approxunion T \sum_{o=1}^O K_o$ continuous decision variables and $L_\approxunion \sum_t \sum_o \sum_k |\Cal{V}_{t,o}^k|$ linear constraints\noteIan{Duplicate information given directly below~\eqref{eq:problem-LP}}, and 
	It contains $L_{\approxunion}T\sum_{o=1}^O K_o$ binary variables and $L_{\approxunion}T\sum_{o=1}^O K_o$ mixed-integer constraints \eqref{mp1:collision-avoidance}.
	Compared with scenario problem \eqref{eq:problem_scenario}, which involves $LTO$ binary variables and $L TON$ mixed-integer constraints \eqref{opt2:chance-constraint}, problem \eqref{eq:problem_motionplan_overapprox} requires more binary decision variables because $L$ is similar to $L_\approxunion$ and $K_o > 1$ if the distribution is multimodal, but it significantly reduces the number of mixed-integer constraints as the number of clusters is much smaller than the number of samples, i.e., $\sum_o K_o \ll ON$. 
	{
		\begin{remark}
			Recall that we use a linear time-varying model for the EV in \eqref{opt1:model} and \eqref{opt1:output}. To handle a nonlinear dynamic model, we can linearize it at each time step to obtain a linear time-varying model, and use the same motion planning formulation as \eqref{eq:problem_motionplan_overapprox}. In fact, this stepwise linearization is a common approach in vehicle control to handle nonlinear dynamics \cite{falcone_predictive_2007}. For OVs, any dynamic model can be used because we only use samples of their predicted trajectories to construct the polytopic constraints \eqref{mp1:collision-avoidance}.
			%			n problem \eqref{eq:problem_motionplan_overapprox}, we consider a linear model for \eqref{opt1:model} and \eqref{opt1:output} so that it consists of linear constraints. Using a linear dynamical model of an autonomous vehicle is common in the literature of high-level motion planning \cite{pek_fail-safe_2020}. However, considering a nonlinear model in \eqref{eq:problem_motionplan_overapprox} is also possible using methods of nonlinear model predictive control \cite{gros_nmpc_2020}, although they require additional computation time. 
		\end{remark}
	}

	We now present the main result that generalizes a feasible solution of \eqref{eq:problem_motionplan_overapprox} to that of chance-constrained problem \eqref{eq:problem_motionplan_prob}.
	{
		\begin{theorem}\label{theorem1}
			Suppose that Assumption~\ref{assume:clustering} holds true. Given a risk level $\epsilon \in (0,1)$ and a confidence bound $\beta \in (0,1)$, let $\epsilon_{o}^k\in (0,1)$ and $\beta_{o}^k\in (0,1)$ be chosen such that 
			\begin{align}\label{eq:theorem_eps_beta}
			\sum_{o=1}^O \sum_{k=1}^{K_o} \epsilon_{o}^k = \epsilon, && 	\sum_{o=1}^O \sum_{k=1}^{K_o} \beta_{o}^k  = \beta.
			\end{align} Let the number of samples $N_{o}^k$ satisfy \eqref{eq:N-condition} with $\epsilon_{o}^k, \beta_{o}^k, n_c = L_{\approxunion}T$ and $n_b = 0$. Then, a feasible solution of \eqref{eq:problem_motionplan_overapprox} is a feasible solution of \eqref{eq:problem_motionplan_prob} with probability at least $1-\beta$.
		\end{theorem}
	}

	\begin{proof}
		{
			Let $\Vc{y}^* = (y_1^*, \ldots, y_T^*)$ denote the output corresponding to \eqref{opt1:model} and \eqref{opt1:output} from a feasible solution of \eqref{eq:problem_motionplan_overapprox}, and $\Bs{\approxunion}_{o}^{*k}$ denote the optimal overapproximation of \eqref{eq:problem-LP}. For each $t$ and $o$, let us partition the set of samples into $\Cal{S}_{t,o}^1, \ldots, \Cal{S}_{t,o}^{K_o}$ where $\Cal{S}_{t,o}^k := \{\Rm{X}_{t,o}\in \R^{n_\Rm{X}}: g(\Rm{X}_{t,o}) = k\}$. Such a partition is given by Assumption~\ref{assume:clustering}. Let $\Bs{\Cal{S}}_{o}^k := \Cal{S}_{1,o}^k \times \ldots \times \Cal{S}_{T,o}^{k}$.}
		%	$\Vc{X}_{o}:= (\Rm{X}_{1,o},\ldots, \Rm{X}_{T,o}), \Vc{X}:= (\Rm{X}_{t,o}: t\in \Z_{1:T}, o\in \Z_{1:O}),$ and 
		
		{
			To prove the theorem, we want to show $\Pr^{N} \left( {V}(\Vc{y}^*) \leq \epsilon   \right) \geq 1- \beta$ where $N = \max N_o^k$ and ${V}(\Vc{y}^*) := \Pr\left(\Vc{X}: \exists t, o\text{ such that }y_t^* \in \Cal{O}(\Rm{X}_{t,o}) \right)$. Here, $\Pr^N$ is the probability measure of $N$ samples $\Vc{X}^{(i)}$.
			To do this, we use $V(\Bs{\approxunion}_{o}^{*k}):= \Pr (\Vc{X}_o\in \Bs{\Cal{S}}_o^k: \exists t \text{ such that } \Cal{O}(\Rm{X}_{t,o})\nsubseteq  \approxunion_{t,o}^{*k})$, and bound ${V}(\Vc{y}^*)$  in terms of $V(\Bs{\approxunion}_o^{*k})$ so that eventually we can bound $\Pr^N(V(\Vc{y}^*)\leq \epsilon)$ in terms of $\Pr^N(V(\Bs{\approxunion}_o^{*k})\leq \epsilon_o^k)$.}
		
		{
			Since $\Vc{y}^*$ is a feasible solution of \eqref{eq:problem_motionplan_overapprox}, we have $y_t^* \notin \approxunion_{t,o}^{*k}$ for any $t,o,$ and $k$. Thus, $\Cal{O}(\Rm{X}_{t,o}) \subseteq \approxunion_{t,o}^{*}$ implies $y_t^* \notin \Cal{O}(\Rm{X}_{t,o})$, which leads to {\small$1- {V}(\Vc{y}^*)=
				\Pr \left( \Vc{X}: \bigwedge_{t,o} y_t^* \notin \Cal{O}(\Rm{X}_{t,o})\right)
				\geq \Pr \left(\bigwedge_{t,o} \Cal{O}(\Rm{X}_{t,o}) \subseteq \approxunion_{t,o} ^*\right) = 1- \Pr \left(\bigvee_{t,o} \Cal{O}(\Rm{X}_{t,o}) \nsubseteq \approxunion_{t,o} ^*\right)
				\geq 1-\sum_o \Pr \left( \bigvee_t \Cal{O}(\Rm{X}_{t,o}) \nsubseteq \approxunion_{t,o} ^*\right)
				= 1-\sum_{o,k}  \Pr \left( \Vc{X}_o \in \Bs{\Cal{S}}_o^k:\bigvee_t \Cal{O}(\Rm{X}_{t,o}) \nsubseteq \approxunion_{t,o} ^{*k}\right) = 1- \sum_{o,k} V(\Bs{\approxunion}_o^{*k}).$}
			The second inequality is by Boole's inequality, i.e., $\Pr(\bigvee_i A_i)\leq \sum_i \Pr(A_i)$. Thus, ${V}(\Vc{y}^*) \leq \sum_{o,k} V(\Bs{\approxunion}_o^{*k}).$}
		
		{
			Since $N_o^k$ satisfies \eqref{eq:N-condition}, by Lemma~\ref{lemma:scenario} $\Pr^N(V(\Bs{\approxunion}_{o}^{*k}) \leq \epsilon_o^k)\geq 1-\beta_o^k$. This allows us to conclude the proof because {\small $\Pr^{N} \left( {V}(\Vc{y}^*) \leq \epsilon   \right) 
				\geq \Pr^{N} \left( \sum_{o,k}V(\Bs{\approxunion}_o^{*k}) \leq \epsilon   \right)  \geq \Pr^N(\bigwedge_{o,k} V(\Bs{\approxunion}_{o}^{*k})\leq \epsilon_o^k) \geq 1-\sum_{o,k} \Pr^N(V(\Bs{\approxunion}_o^{*k})\leq \epsilon_{o}^k) \geq 1 - \beta$.}}
	\end{proof}
	
	Theorem~\ref{theorem1} illustrates that if we have $N_o^k$ samples to compute $\Bs{\approxunion}_{o}^k$ for each OV $o$ and cluster $k$, then any feasible solution of our approach (i.e., \eqref{eq:problem-LP} and \eqref{eq:problem_motionplan_overapprox}) is a feasible solution to \eqref{eq:problem_motionplan_prob} with probability at least $1-\beta$. 
	
	{In Theorem~\ref{theorem1}, $\epsilon_o^k$ and $\beta_o^k$ that satisfies \eqref{eq:theorem_eps_beta} can be chosen based on problem-specific knowledge. For example, in the case study in Section~\ref{sec:intersection}, the probability $p(g(\Rm{X}_{t,o})=k)$ is known.  In that case we can set $\epsilon_o^k$ for each OV to be proportional to the inverse of $p(g(\Rm{X}_{t,o})=k)$ so that a less likely cluster is assigned a higher risk and thus requires fewer samples. Another choice is to allocate $\epsilon$ and $\beta$ uniformly over clusters: $\epsilon_{o}^k = \epsilon / \sum_{o=1}^{O} K_o$ and $\beta_o^k = \beta / \sum_{o=1}^O {K_o} $.}
	
	\begin{example}\label{ex:preprocessing}
		Recall Example~\ref{ex:scenario-approach}. Now, we separate the samples into two clusters and compute $\Cal{C}^1$ and $\Cal{C}^2$ by \eqref{eq:problem-LP} given predetermined vectors $\bar{A}^1 = \bar{A}^2 = (1, -1)^\intercal$. Then \eqref{eq:problem_motionplan_overapprox} becomes the problem that minimizes $|y|$ subject to $y\notin \approxunion^1$ and $y \notin \approxunion^2$. 
		%	\begin{subequations}\label{eq:problem_example2}
		%		\begin{align}
		%%		&\min_{y, \Vc{z}\in \Cal{Z}}  && |y|\\
		%& y + M \left(1-z_1^k\right) \geq \Rm{X}^{(i)} + 0.1,  && i \in I^k,\\
		%&-y + M \left(1-z_2^k\right) \geq -\Rm{X}^{(i)} + 0.1,  && i \in I^k,
		%		\end{align}
		%	\end{subequations}
		%Here, $\Cal{Z} = \{ (z_1^1, z_2^1, z_1^2, z_2^2)\in \{0,1\}^4: z_1^1 + z_2^1 \geq 1, z_1^2 + z_2^2 \geq 1\}$. 
		By Theorem~\ref{theorem1} with the uniform allocation of $\epsilon$ and $\beta$, if $N^1 = N^2 \geq 400.6$, any feasible solution of the problem ensures $\Pr (y\notin \Cal{O}(\Rm{X})) \geq 1-0.05$ with probability at least $1-0.01$. The optimal cost is $0$ because $y=0$ is feasible, as illustrated in Fig.~\ref{fig:scenario-example}(b). Compared with Example~\ref{ex:scenario-approach}, our approach yields a less conservative solution.
	\end{example}
	
	%Example~\ref{ex:preprocessing} illustrates that our approach is less conservative. 
	In the next section, we will show that even for a higher dimensional practical example, our approach is less conservative and computationally more efficient than the standard scenario approach without preprocessing.
	
	%%%%%%%%%%%%%%%%%%%%%%%%%%%%%%%%%%%%%%%%%%%%%%%%%%%%%%%%%%%%%%%%%%%%%%%%%%%%%%%%
	\section{Simulation Result}\label{sec:simulation}
	
	In this section, we compare the performance of our approach to that of the basic scenario approach via simulations.\footnote{All computations have been performed on a computer running macOS 10.15 with a 3.5 GHz Intel i5 CPU and 8 GB memory. Our implementation is written for MATLAB using CPLEX 12.10 as an optimization solver.} We model the EV dynamics as double integrators with speed and input constraints as in the case studies of \cite{sessa_uncertainty_2018}. In our first case study, we impose multimodal uncertainties on the acceleration of the OV to generate trajectory predictions. In the second case study, we integrate our approach with the state-of-the-art forecasting neural network Trajectron++ \cite{salzmann_trajectron_2021}.	

	\begin{figure}[t]
		\centering
		\includegraphics[ width =0.9\linewidth]{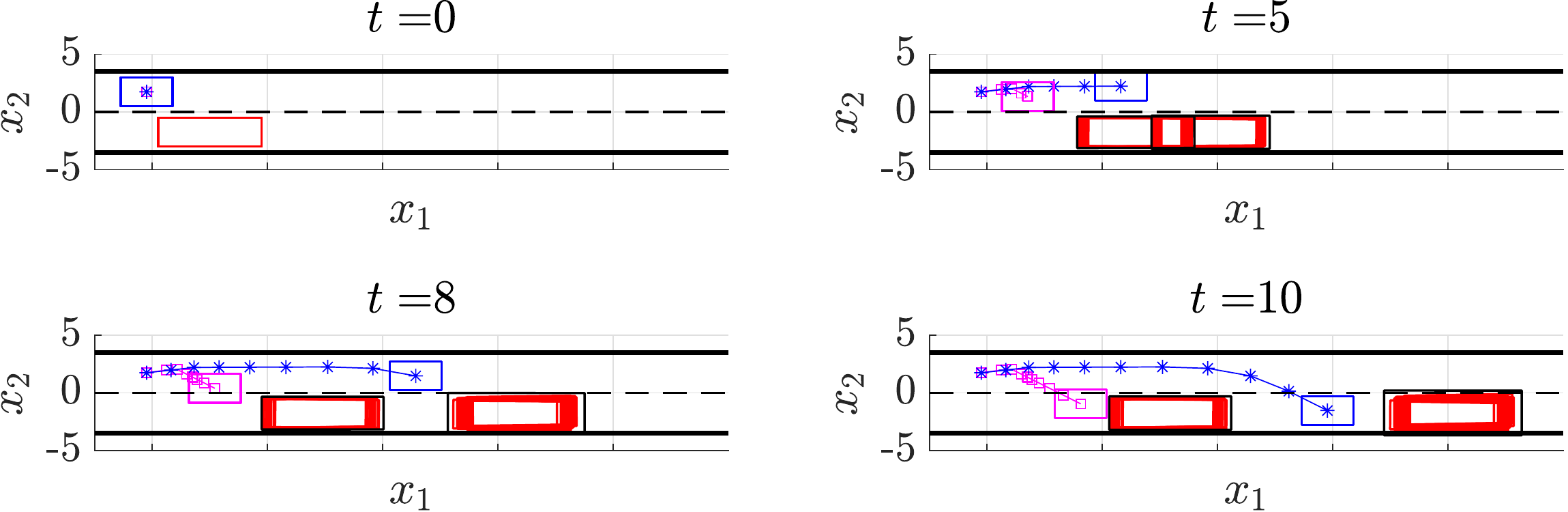}
		\caption{Lane change scenario of the EV (magenta/blue boxes). Predicted trajectories of the obstacle truck (red boxes) show the bimodality of decelerating and accelerating. Our approach (blue) yields a less conservative result than the basic scenario approach (magenta). }
		\label{fig:lane_change}
	\end{figure}
	\subsection{Case study 1: Lane change}\label{sec:lane_change_simulation}
	In Fig.~\ref{fig:lane_change}, the EV intends to change lane, and the OV is predicted to decelerate or accelerate. The OV's predicted potential obstacle sets are shown in red, and the EV's position resulting from our approach and the basic scenario approach is shown in blue and magenta, respectively. The objective here is to progress furthest to the right while achieving the bottom lane. With $\epsilon = 0.05, \beta = 10^{-3}, L_\approxunion = 4, T = 10, O = 1, K_1 = 2$ and uniform allocation of $\epsilon$ and $\beta$, we use $N_1^1 = N_1^2 = 2964$ trajectories (in total $N = 5928$) based on Theorem~\ref{theorem1} for our scheme. As reported in Table~\ref{table:comparison}, the basic scenario approach requires fewer samples, but our approach is more than 20 times faster because the number of samples $N$ affects only the computation time of the linear program \eqref{eq:problem-LP}, not that of the mixed-integer program \eqref{eq:problem_motionplan_overapprox}. The reported computation time does not include sample generation time. In our approach, clustering, \eqref{eq:problem-LP}, and \eqref{eq:problem_motionplan_overapprox} take $0.06\,$s, $0.02\,$s, and $0.1\,$s, respectively (summing to $0.18\,$s). Also, our approach is less conservative in the sense that it yields a better objective cost. 
	%, which is encoded as minimizing the negative of the maximum of $x_1$ and the absolute difference between the center of the bottom lane and $x_2$ at time step $T$. 
	Indeed, Fig.~\ref{fig:lane_change} shows that the EV slows down to change lanes in the basic scenario approach, whereas in our approach it neatly fits itself between the two possible OV behaviors. For the optimal output $\Vc{y}^*$ of our approach, we checked whether $y^*_t \notin \Cal{O}(\Rm{X}_{t,o}^{(i)})$ for all $t\in \Z_{1:T}$ for $10^5$ newly predicted trajectories. The empirical violation---the fraction of sampled cases which demonstrate violations---is higher in our approach because it enables less conservative motions, but is still within the specified risk level of $\epsilon = 0.05$. \revised{In some cases (e.g., when the EV initial velocity is large), our approach fails in finding a lane-changing trajectory, but still guarantees safety at the risk level of $\epsilon$ with high confidence.}
	%We have not tried to optimize the computation performance of our approach. By parallelizing the computation of overapproximations in \eqref{eq:problem-LP} or using computationally more efficient programming languages, such as Python or C, we could further improve the performance to enable real-time implementation. 
	
	\begin{table}[t!]
		\caption{}
		\begin{center}
			\begin{tabular}{ |l|r|r| } 
				\hline
				& \multicolumn{1}{|c|}{Scenario \eqref{eq:problem_scenario}}& \multicolumn{1}{|c|}{Our approach \eqref{eq:problem-LP}+\eqref{eq:problem_motionplan_overapprox}}\\ 
				\hline  
				$N$ &  1,706 & {5,928}\\
				Computation time (s) & {4.20} & {0.18}\\ 
				Minimum cost & {-2.08} & {-6.79}\\
				Empirical violation & {0.0025} & {0.0334} \\
				\hline
			\end{tabular}
		\end{center}\label{table:comparison}
		
	\end{table}

	\begin{figure}[t]
		\centering
		\includegraphics[ width = 0.7\linewidth]{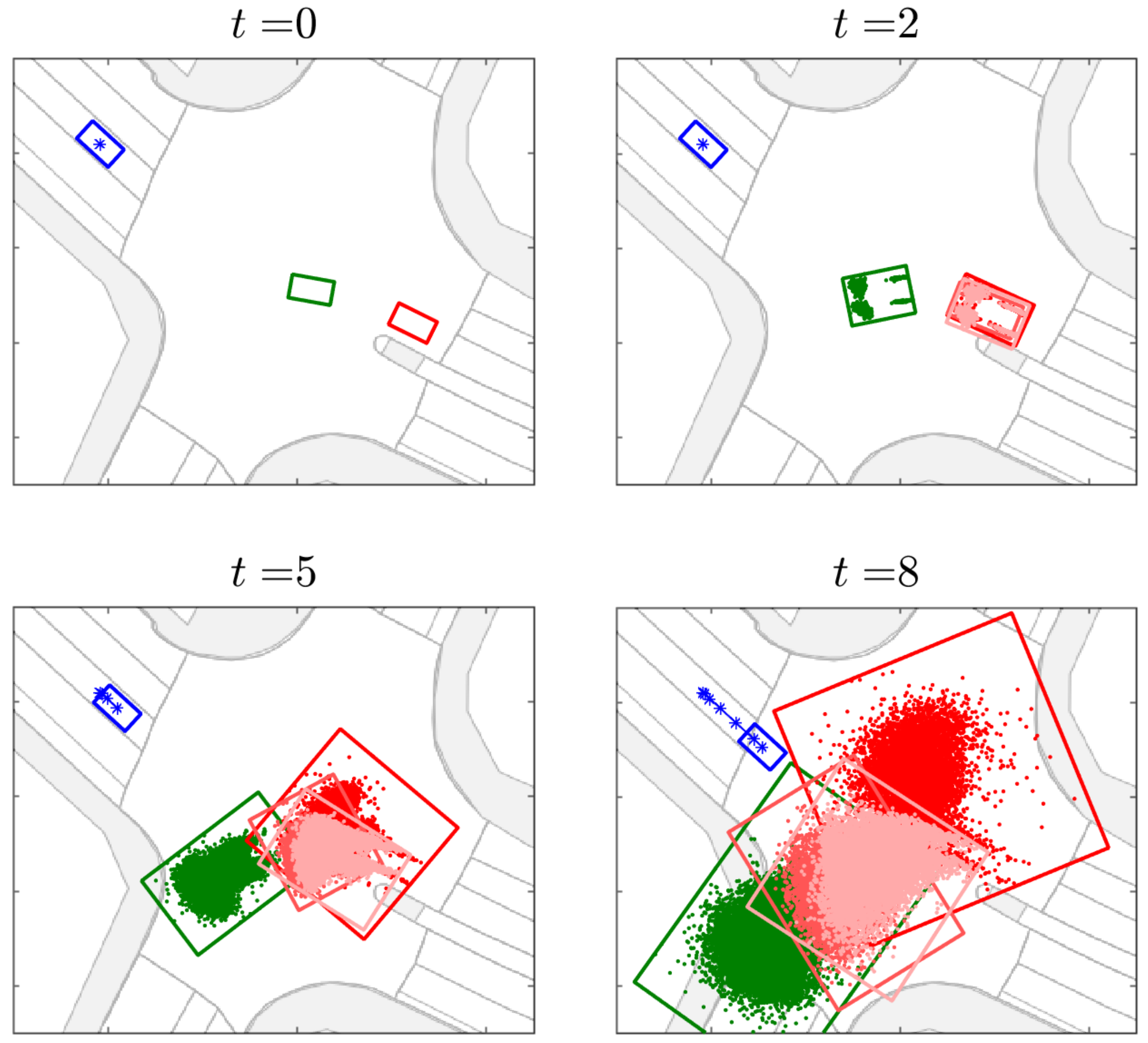}
		\caption{{Predictions of two OVs (red and green boxes at $t=0$) obtained from Trajectron++ \cite{salzmann_trajectron_2021} and motion of the EV (blue boxes) computed by our approach, on an testing example of the nuScenes dataset \cite{caesar2020nuscenes}.}}
		\label{fig:nuscene_crossing}
	\end{figure}

	{\subsection{Case study 2: Intersection on the nuScenes dataset}\label{sec:intersection}
		We train the forecasting neural network {Trajectron++} \cite{salzmann_trajectron_2021} on the nuScenes dataset \cite{caesar2020nuscenes} using Compute Canada\footnote{\url{https://www.computecanada.ca/}} and integrate our motion planning approach with its testing results. In particular, we modify Trajectron++ such that at test time, it outputs $N$ samples of $(\Vc{X}^{(i)}, \Vc{z}^{(i)})$ where $\Vc{z} = (z_o, o\in \Z_{1: O})$ is a discrete latent variable. Since the latent variable encodes high-level intent, we use $\Vc{z}^{(i)}$ to define the clustering function $g$ by setting $i\in I_o^k$ if ${z}_o^{(i)} = k$. To reduce the number of clusters, we select a subset $\tilde{\Vc z}_o$ of latent variables for each $o$ that satisfy $p({z}_o|\phi)\geq 0.1$, reassign all forecast trajectories for the latent variables which were discarded to one of $\tilde{\Vc z}_o$, and recompute $p(z_o|\phi)$ such that $\sum_{z_o\in \tilde{\Vc z}_o} p(z_o|\phi) = 1$.}
	
	%	 based on $p(\Vc{X}_{1:T}|\phi) = \sum_{\Vc{z}} p(\Vc{X}_{1:T}| \phi, \Vc{z}) p(\Vc{z}|\phi)$
	% given $\phi$, $\Vc{z}^{(i)} \sim p(\Vc{z}| \phi)$ and $\Vc{X}_{1:T}^{(i)} \sim p(\Vc{X}_{1:T}| \Vc{z}, \phi)$. 
	%We  Since each discrete latent variable represents a cluster of future trajectories, 
	%Given a training dataset $\Cal{D}=(\Vc{X}_{1:T}^i, \phi^i)$, {Trajectron++} learns the distribution over discrete latent variables $p(\Vc{z}|\Cal{D})$ and conditional distribution over future trajectories $p(\Vc{X}_{1:T}| \Vc{z}, \Cal{D})$. 

	{In Fig.~\ref{fig:nuscene_crossing}, the EV (blue) approaches the intersection, and the two OVs (red and green at $t=0$) are predicted to turn. The red OV exhibits three modes (red, dark pink, and light pink); although the dark and light pink predictions are not separable in the 2-dimensional space, we refer to them as different modes because they correspond to different latent variables. The green OV exhibits one mode. 	With $\epsilon = 0.1, \beta=10^{-3}, L_\approxunion = 4, T = 8, O=2, K_1 = 3, K_2 = 1$, we choose $\epsilon_2^1 = \epsilon/9.14$ and $\beta_2^1 = \beta/9.14$ and assign the rest to $\epsilon_1^k$ and $\beta_1^k$ such that they are proportional to $1/p({z}_1 = k|\phi)$. The purpose of this allocation is to match the number of samples for each OV, i.e, $\sum_{k=1}^3 N_1^k \approx N_2^1$, because a prediction from Trajectron++ includes trajectories of all agents to account for their interaction. This choice results in extracting $\sum_{k=1}^3 N_1^k = 5941$ and $N_2^1 = 5831$ predictions according to Theorem~\ref{theorem1}. We compute rectangular overapproximations (red, dark pink, light pink, and green boxes) that contain all of the obstacle sets (whose vertexes are represented by the dots of the same colors). The objective of the EV is to progress furthest in its longitudinal direction while minimizing its lateral distance and velocity at the end of the planning horizon, so that the EV comes to an halt even if the red OV does not commit to the turn after $T$ time steps. In $0.14\,$s, our approach computes a EV trajectory for the $4\,$s horizon that avoids collisions with the overapproximations. Zero empirical violations were detected in $10^5$ new predictions.	Scenario problem \eqref{eq:problem_scenario}  infeasible for this example, which demonstrates that it is more conservative.}

	{We conclude this section with several observations.  First, although the computation complexity of \eqref{eq:problem_motionplan_overapprox}  grows exponentially with the number $O$ of OVs because the number of binary variables increases with $O$, we observed that for a moderate number of OVs (e.g., $O=9$, each with $K_o=1$), the computation can be done within an alloted time of $0.5\,$s. 
		The main bottleneck of applying our approach to multiple-OV scenarios has been the increasing number of samples $N$ required by Theorem~\ref{theorem1} to ensure probabilistic safety. Our approach thus requires prediction methods that can efficiently provide a large number of samples. 
		Second, the conservativeness of our approach depends on the quality of the overapproximations based on \eqref{eq:problem-LP}. We observed that with a rectangular representation (i.e., $L_\approxunion = 4$), our approach was always less conservative than the basic scenario approach. Lastly, state-of-the-art predictions are highly variable and often prevent the EV from progressing. For example, in Fig.~\ref{fig:nuscene_crossing} the EV cannot cross the intersection as no gap exists between the modes of the red OV. Our approach can take advantage of future improvements in prediction performance because it can efficiently handle multimodality (as shown in Section~\ref{sec:lane_change_simulation}).}

	\section{Conclusion}
	We have presented a motion planning approach that ensures probabilistic safety in the presence of multimodal uncertainties. 
	%\noteIan{What about your uncertainty model is specific to DNNs? - \revised{Not sure about the question.. did I answer your question with the rephrased sentence?}  From my reading, you do not need to use a DNN to generate the samples (in fact, your experiment did not).  So you have created an algorithm which ensures safety for any distribution from which you can sample.  That happens to include the sample-generating DNNs that you use as motivation, but could also be Gaussian processes, MDPs, or pseudo-random closed-form models (such as used in your examples).} 
	%Motivated by autonomous driving examples where DNNs are used to predict future trajectories of other vehicles, we have focused on uncertainties that have a multimodal distribution. 
	We have formulated the motion planning problem as a chance-constrained problem and presented an efficient, sampling-based approach to obtain a high-confidence solution. 
	%Our approach includes two subproblems: first, we cluster all samples into the same number of groups as the number of modes in the multimodal distribution and compute an overapproximate set for each group, and next, we solve a robust motion planning problem against the overapproximate sets. 
	In particular, we have characterized the number of samples required to guarantee probabilistic safety with high confidence. We have validated on a real-world dataset that our approach is less conservative and computationally more efficient than a conventional scenario approach. {Since our approach works for any multimodal distribution from which we can sample, it could be integrated with, for example, prediction models based on Gaussian processes or Markov Decision Processes.  We are currently implementing a receding horizon version of our approach to achieve closed-loop control in dynamic environments.}
	%As future work, we plan to relax the assumptions on the accuracy of DNNs to account for training errors and out-of-distribution scenarios. Also, we plan to implement our proposed control method on a realistic vehicle simulator.
	
	%\addtolength{\textheight}{-3cm}   
	
	\bibliographystyle{IEEEtran}
	\bibliography{IEEEabrv}

\begin{thebibliography}{10}
\providecommand{\url}[1]{#1}
\csname url@samestyle\endcsname
\providecommand{\newblock}{\relax}
\providecommand{\bibinfo}[2]{#2}
\providecommand{\BIBentrySTDinterwordspacing}{\spaceskip=0pt\relax}
\providecommand{\BIBentryALTinterwordstretchfactor}{4}
\providecommand{\BIBentryALTinterwordspacing}{\spaceskip=\fontdimen2\font plus
\BIBentryALTinterwordstretchfactor\fontdimen3\font minus
  \fontdimen4\font\relax}
\providecommand{\BIBforeignlanguage}[2]{{%
\expandafter\ifx\csname l@#1\endcsname\relax
\typeout{** WARNING: IEEEtran.bst: No hyphenation pattern has been}%
\typeout{** loaded for the language `#1'. Using the pattern for}%
\typeout{** the default language instead.}%
\else
\language=\csname l@#1\endcsname
\fi
#2}}
\providecommand{\BIBdecl}{\relax}
\BIBdecl

\bibitem{fisac_general_2019}
J.~F. Fisac \emph{et~al.}, ``A general safety framework for learning-based
  control in uncertain robotic systems,'' \emph{IEEE Trans. Autom. Control},
  vol.~64, no.~7, pp. 2737--2752, Jul. 2019.

\bibitem{wabersich_probabilistic_2021}
K.~P. Wabersich \emph{et~al.}, ``Probabilistic model predictive safety
  certification for learning-based control,'' \emph{IEEE Trans. Autom.
  Control}, Jan. 2021, {Early} Access.

\bibitem{carvalho_automated_2015}
A.~Carvalho \emph{et~al.}, ``Automated driving: {The} role of forecasts and
  uncertainty -- {A} control perspective,'' \emph{European J. Control},
  vol.~24, pp. 14--32, Jul. 2015.

\bibitem{sessa_uncertainty_2018}
P.~G. Sessa \emph{et~al.}, ``From uncertainty data to robust policies for
  temporal logic planning,'' in \emph{Proc. ACM Int. Conf. Hybrid Systems:
  Computation and Control (HSCC)}, Apr. 2018, pp. 157--166.

\bibitem{lefkopoulos_using_2019}
V.~Lefkopoulos and M.~Kamgarpour, ``Using uncertainty data in
  chance-constrained trajectory planning,'' in \emph{Proc. European Control
  Conference (ECC)}, Jun. 2019, pp. 2264--2269.

\bibitem{salzmann_trajectron_2021}
T.~Salzmann \emph{et~al.}, ``Trajectron++: {Dynamically}-feasible trajectory
  forecasting with heterogeneous data,'' \emph{arXiv:2001.03093}, Jan. 2021.

\bibitem{ivanovic_mats_2021}
B.~Ivanovic \emph{et~al.}, ``{MATS}: {An} interpretable trajectory forecasting
  representation for planning and control,'' \emph{arXiv:2009.07517}, Jan.
  2021.

\bibitem{calafiore_distributionally_2006}
G.~C. Calafiore and L.~E. Ghaoui, ``On distributionally robust
  chance-constrained linear programs,'' \emph{Journal of Optimization Theory
  and Applications}, vol. 130, pp. 1--22, Dec. 2006.

\bibitem{hu_chance_2018}
\BIBentryALTinterwordspacing
Z.~Hu \emph{et~al.}, ``Chance constrained programs with mixture
  distributions,'' 2018. [Online]. Available:
  \url{http://www.optimization-online.org/DB_FILE/2018/09/6798.pdf}
\BIBentrySTDinterwordspacing

\bibitem{esfahani_performance_2015}
P.~M. Esfahani \emph{et~al.}, ``Performance bounds for the scenario approach
  and an extension to a class of non-convex programs,'' \emph{IEEE Trans.
  Autom. Control}, vol.~60, no.~1, pp. 46--58, Jan. 2015.

\bibitem{margellos_road_2014}
K.~Margellos \emph{et~al.}, ``On the road between robust optimization and the
  scenario approach for chance constrained optimization problems,'' \emph{IEEE
  Trans. Autom. Control}, vol.~59, no.~8, pp. 2258--2263, Aug. 2014.

\bibitem{caesar2020nuscenes}
H.~Caesar \emph{et~al.}, ``nu{S}cenes: A multimodal dataset for autonomous
  driving,'' \emph{arXiv:1903.11027}, May 2020.

\bibitem{bemporad1999control}
A.~Bemporad and M.~Morari, ``Control of systems integrating logic, dynamics,
  and constraints,'' \emph{Automatica}, vol.~35, no.~3, pp. 407--427, Mar.
  1999.

\bibitem{nalisnick_deep_2018}
E.~Nalisnick \emph{et~al.}, ``Do deep generative models know what they don't
  know?'' in \emph{Int. Conf. Learning Representations (ICLR)}, Feb. 2019.

\bibitem{calafiore_uncertain_2005}
G.~Calafiore and M.~Campi, ``Uncertain convex programs: {R}andomized solutions
  and confidence levels,'' \emph{Mathematical Programming}, vol. 102, no.~1,
  pp. 25--46, Jan. 2005.

\bibitem{bishop2006pattern}
C.~M. Bishop, \emph{Pattern recognition and machine learning}.\hskip 1em plus
  0.5em minus 0.4em\relax Springer, 2006.

\bibitem{falcone_predictive_2007}
P.~Falcone \emph{et~al.}, ``Predictive active steering control for autonomous
  vehicle systems,'' \emph{IEEE Trans. Control Syst. Technol.}, vol.~15, no.~3,
  pp. 566--580, May 2007.

\end{thebibliography}
\end{document}